\documentclass[fleqn,twoside]{article}
\usepackage{espcrc2,mathbbol}

\usepackage{graphicx}

\title{
\vspace{-1.5cm}
\begin{flushright}
\normalsize
{\rm\large HU-EP-05/74}\\
\end{flushright}
\vspace{0.2cm}
Laplacian modes as a filter}

\author{Falk Bruckmann\thanks{
Talk given by FB at the Workshop on Computational
Hadron Physics, Cyprus, September 2005.}\address{
  Instituut-Lorentz for Theoretical Physics,
  University of Leiden,
  2300 RA Leiden,
  The Netherlands},
        Ernst-Michael Ilgenfritz\address{
  Institut f\"ur Physik,
  Humboldt Universit\"at zu Berlin,
  12489 Berlin,
  Germany}}
       
\begin{document}

\begin{abstract}
We compute low-lying eigenmodes of the gauge covariant Laplace operator
on the lattice at finite temperature. For classical configurations
we show how the lowest mode localizes the monopole constituents
inside calorons and that it hops upon changing the boundary conditions.
The latter effect we observe for thermalized backgrounds, too,
analogously to what is known for fermion zero modes.

We propose a new filter for equilibrium configurations which provides
link variables as a truncated sum involving the Laplacian modes.
This method not only reproduces classical structures, but also preserves
the confining potential, even when only a few modes are used.
\end{abstract}

%add
%caloron fermion zero mode
%adjoint rep
%corr. Polyakov loop Laplacian mode
%filtered structures: Polyakov loop (Haar), action

\maketitle

\section{Introduction}

As an analyzing tool we study low-lying eigenmodes of the gauge
covariant Laplace operator 
\begin{eqnarray}
&&\hspace*{-0.7cm}\triangle^{ab}_{xy}\!=\!\!\sum_{\mu=1}^4\!\left[
U_\mu^{ab}(x)\delta_{x+\hat{\mu},y}\!+\!
U_\mu^{\dagger ab}(y)\delta_{x-\hat{\mu},y}\!-\!
2\delta^{ab}\delta_{xy}\right]\nonumber\\
&&\hspace*{-0.7cm}-\triangle\phi_n(x)=\lambda_n\phi_n(x)
\label{eqn_1}
\end{eqnarray}
with $U_\mu(x)$ a given lattice configuration in
the fundamental representation of $SU(2)$.
Compared to fermionic (near) zero modes,
Laplacian modes
%are computationally cheaper because they are space-time scalars and
do not suffer from chirality problems nor doubler problems.
They seem not related directly to topology
(like zero modes are, due to index theorems), 
but it has been observed that they are
sensitive to the location of instantons \cite{bruckmann:01a,deforcrand:01a}. 
The scaling
properties of their localization have been investigated 
recently~\cite{greensite:05}.

Here we will present two ideas concerning Laplacian modes
\cite{bruckmann:05b}. 
The first one is to study what can be learned from
the profile of the modulus of the lowest mode 
(the `ground state probability density')
$|\phi_0(x)|^2$, in particular with changing boundary conditions.
The second idea is to introduce a novel low-pass filter.
It concerns the reconstruction of the gauge background
from relatively few Laplacian modes.
The purpose of both methods is to identify the underlying
infrared degrees of freedom in the gauge field
that are responsible for features like confinement.

\section{Profile of the lowest mode}

\subsection{Caloron backgrounds}

\begin{figure}
\includegraphics[width=0.95\linewidth]{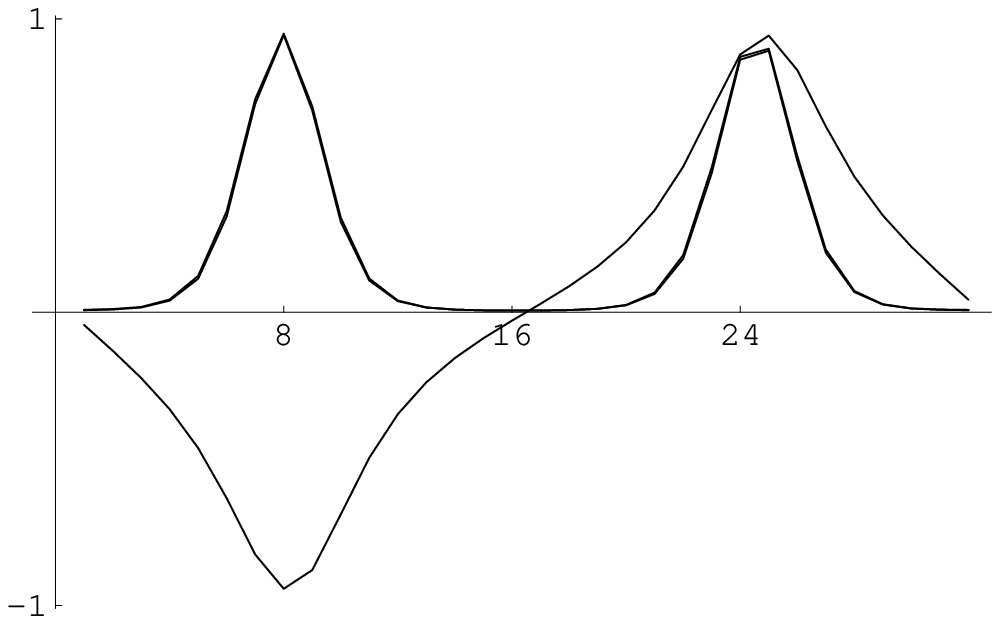}
\includegraphics[width=0.95\linewidth]{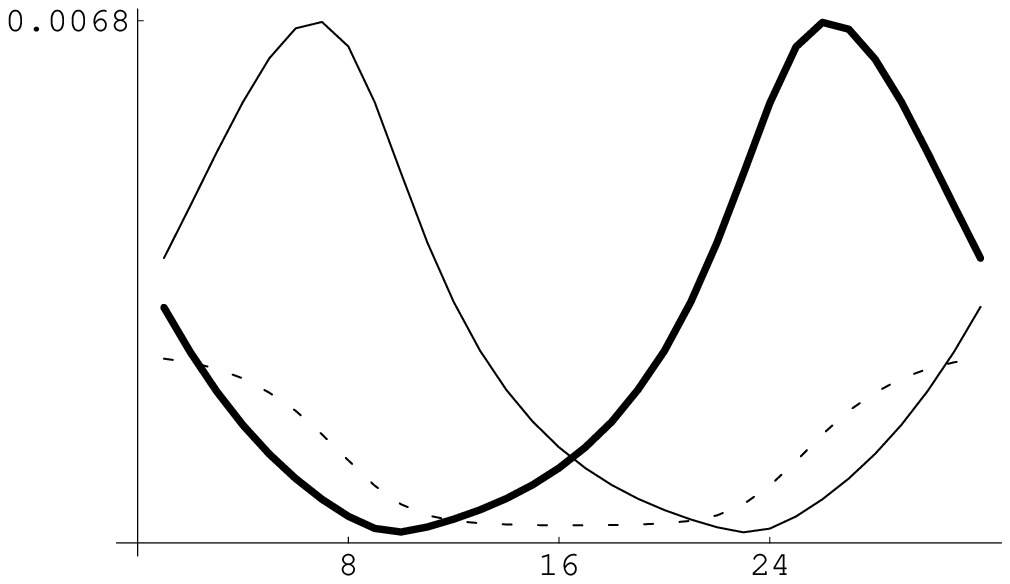}
\caption{Top: action and topological
density (not distinguishable, both multiplied by a factor 500) and
the traced Polyakov loop shown along the line connecting the constituents in a large
caloron on a $32^3\cdot 4$ lattice.
Bottom: profile of the lowest Laplacian mode in that background
with periodic (boldface), antiperiodic (thin) and
intermediate (dashed) boundary conditions, respectively.}
\label{fig_cal_scal_mode}
\end{figure}

As a testing ground for the Laplacian modes we first investigate 
a caloron of maximally nontrivial holonomy 
with its monopole constituents \cite{kraan:98a,lee:98b} 
put on a $32^3\cdot 4$ lattice\footnote{
This is done by calculating links from 
the continuum gauge field, followed by a few steps of cooling.}.
 Fig.\ \ref{fig_cal_scal_mode} (top) shows the action 
density along the line connecting these monopoles, which are
clearly visible as two selfdual (and almost perfectly static) lumps.
As another signal the Polyakov loop goes through $\Eins_2$ and
$-\Eins_2$ -- which amounts to a local symmetry restoration -- at the
monopole cores.

The lowest-lying mode of the Laplacian in this background is shown in Fig.\
\ref{fig_cal_scal_mode} (bottom) \cite{bruckmann:05b}. 
One can see that the presence of monopoles
is reflected by a maximum resp.\ a minimum in the profile of the
lowest mode (although slightly shifted).
In addition, we have depicted the lowest-lying Laplacian mode with
antiperiodic boundary conditions in Euclidean time.
In that mode the role of the monopoles is interchanged,
which can be understood from a symmetry of the caloron.
Hence the lowest-lying Laplacian mode `hops'
as the result of changing the boundary conditions.
It behaves similar to the caloron fermion zero mode \cite{garciaperez:99c}
in the context of which complex boundary conditions were introduced first.
Also for the Laplacian modes we now allow for a general phase in the boundary condition
\begin{equation}
\phi(x_4+N_4)=\exp(2\pi i\zeta)\phi(x_4)\,.
\end{equation}
The angle $\zeta$ can be restricted to $\zeta\in[0,1/2]$ 
since the remaining $\zeta$'s can be mapped to this
interval by charge conjugation.
For the intermediate case $\zeta=1/4$,
i.e.\ halfway between periodic and antiperiodic boundary conditions,
we find a valley in the mode
profile extending between the monopoles (dashed curve in Fig.\
\ref{fig_cal_scal_mode} (bottom)).

The Laplacian eigenmodes are stationary points of the functional
$T[U_\mu;\phi]=\sum_x t(x)$ with
\begin{eqnarray}
t(x)=\sum _\mu |U_\mu^{ab}(x)\phi^b(x+\hat{\mu})-\phi^a(x)|^2
\end{eqnarray}
the lattice analogue of the square of the covariant derivative.
This kinetic term (in the notion of Higgs models) represents the local
distribution of the remaining `energy' when the mode $\phi$ is best adapted to
the background $U_\mu$. Here we view the quantity $t(x)$ as another observable
that is sensitive to structures in the lattice configuration under
consideration.

\begin{figure}
\includegraphics[width=0.95\linewidth]{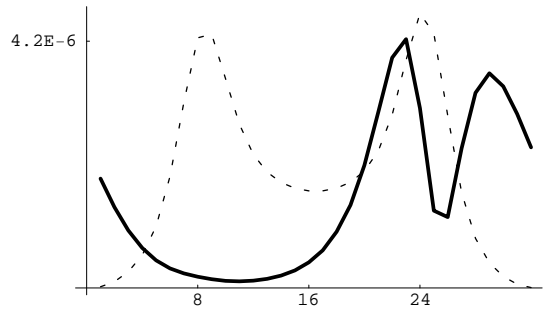}
\caption{The kinetic term for the caloron of Fig.\ \protect\ref{fig_cal_scal_mode} with
periodic (boldface) and intermediate (dashed, multiplied by 8) boundary conditions.}
\label{fig_cal_kin}
\end{figure}

Fig.\ \ref{fig_cal_kin} shows the kinetic term for the caloron and its lowest mode
with different boundary conditions. With the periodic mode the kinetic term
reveals the monopole by a volcano-like structure, i.e.\ a maximum with a dip in
its center. Like the mode itself the kinetic term hops
to the complementary monopole when changing boundary conditions to
antiperiodic (not shown).
For the lowest mode with intermediate boundary conditions {\em both}
monopoles are detected by {\em maxima} in this observable (dashed curve in Fig.\
\ref{fig_cal_kin}).

\subsection{Thermalized backgrounds}
\label{sect_2.2}

\begin{figure*}
\includegraphics[width=0.32\linewidth]
{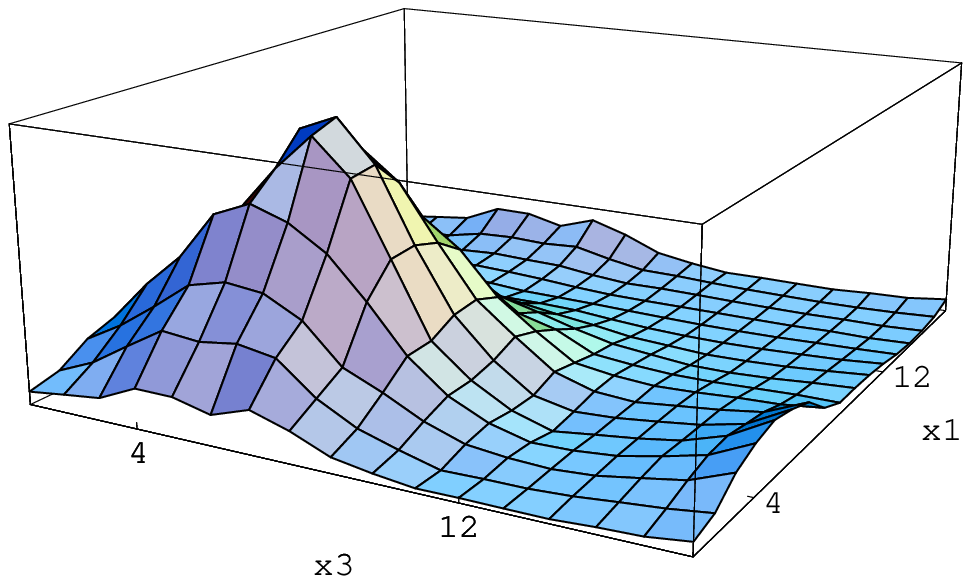}
\includegraphics[width=0.32\linewidth]
{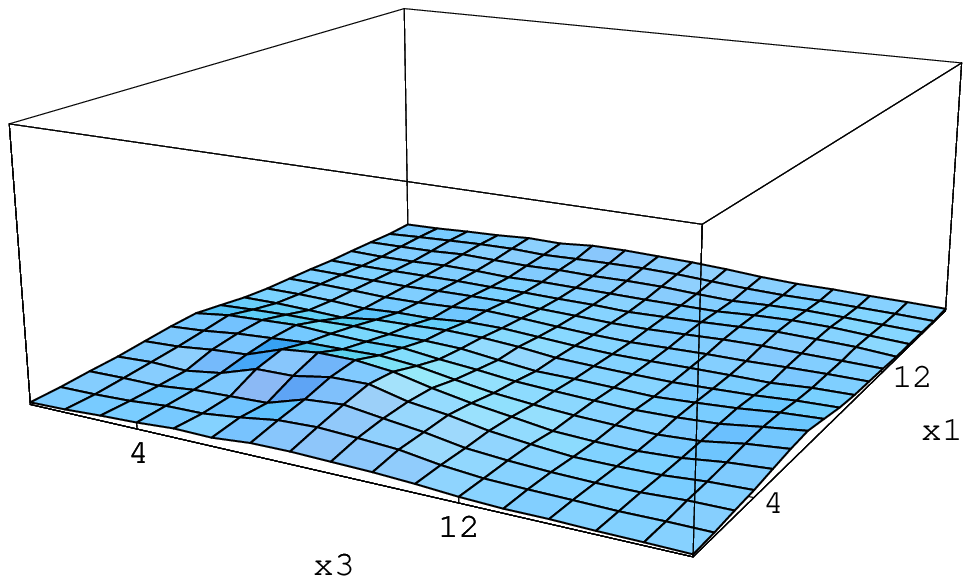}
\includegraphics[width=0.32\linewidth]
{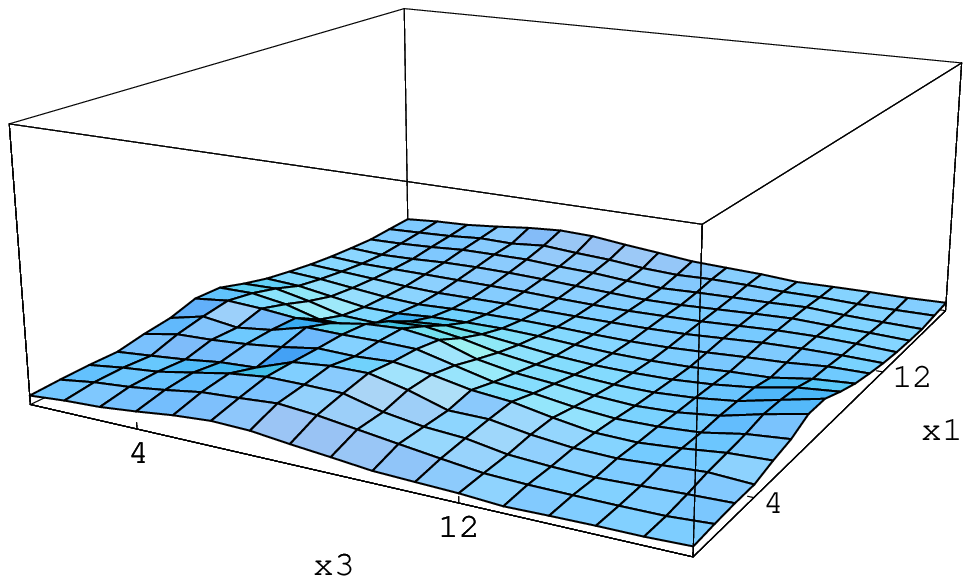}\\
\includegraphics[width=0.32\linewidth]
{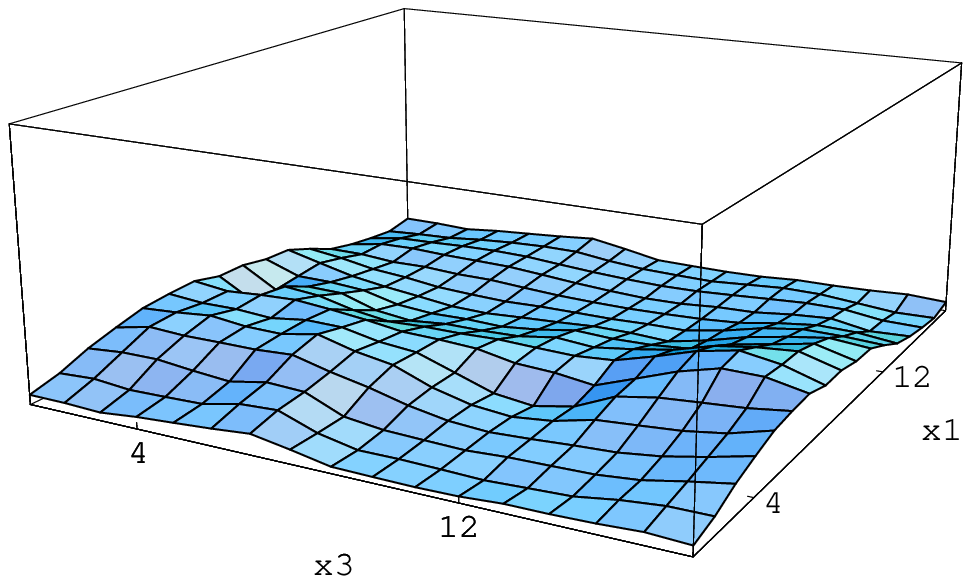}
\includegraphics[width=0.32\linewidth]
{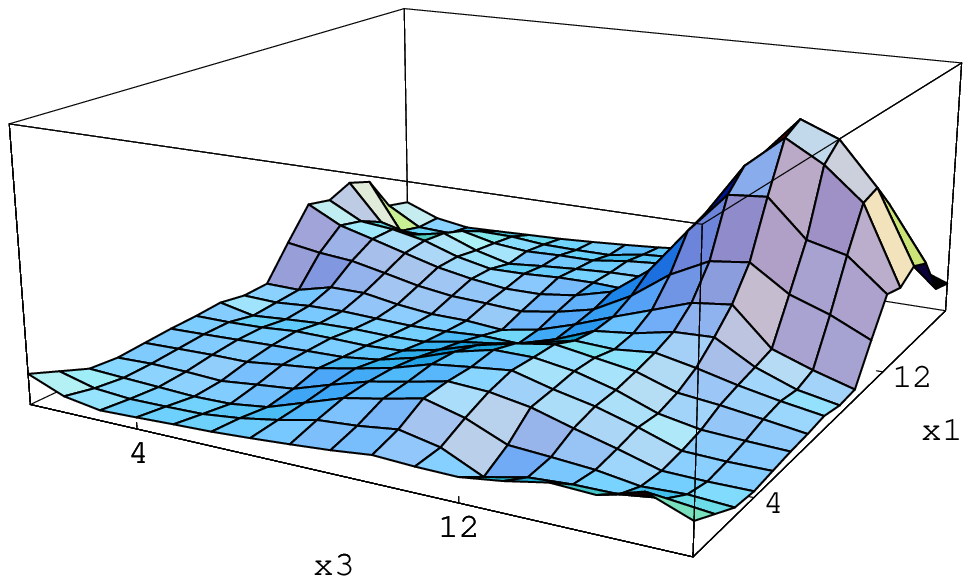}
\includegraphics[width=0.32\linewidth]
{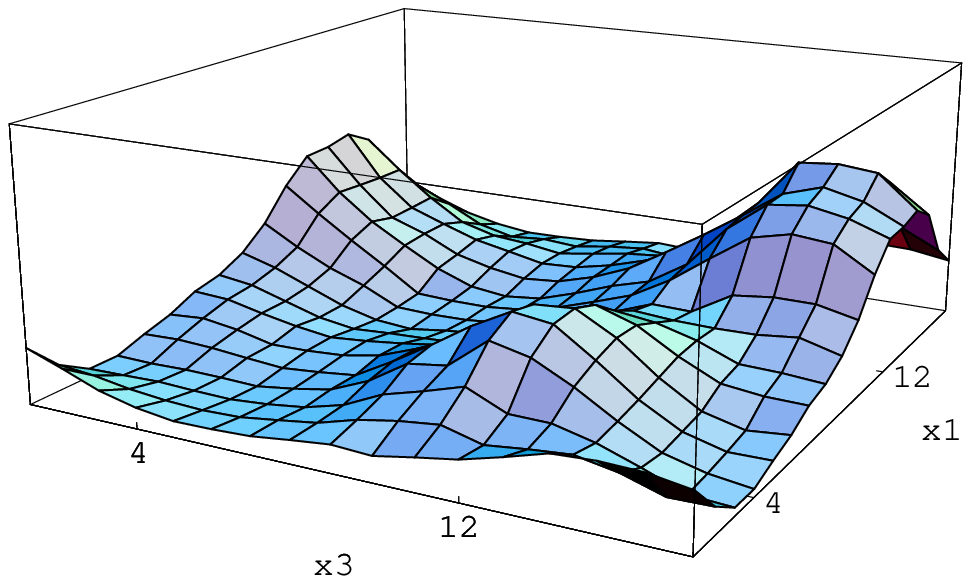}\\
\includegraphics[width=0.32\linewidth]
{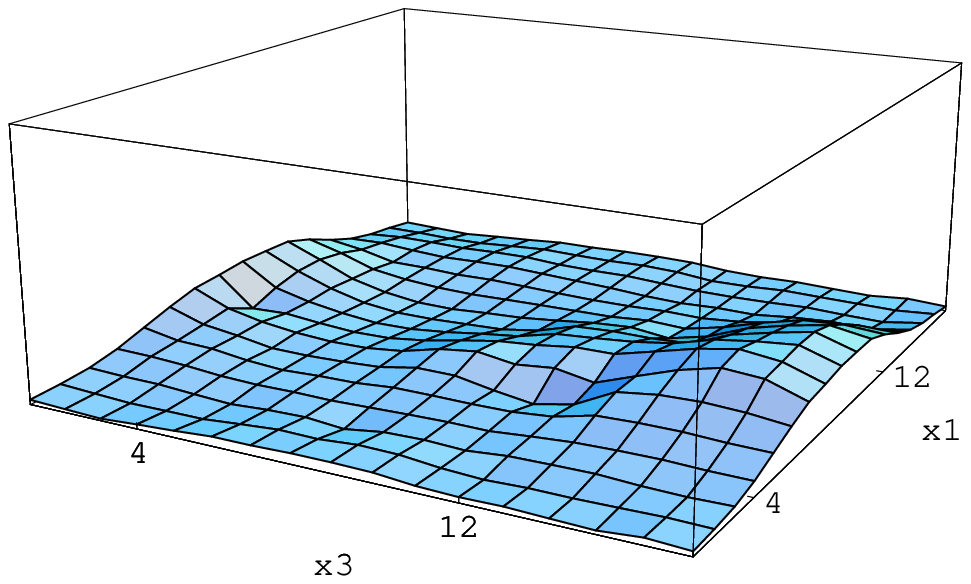}
\includegraphics[width=0.32\linewidth]
{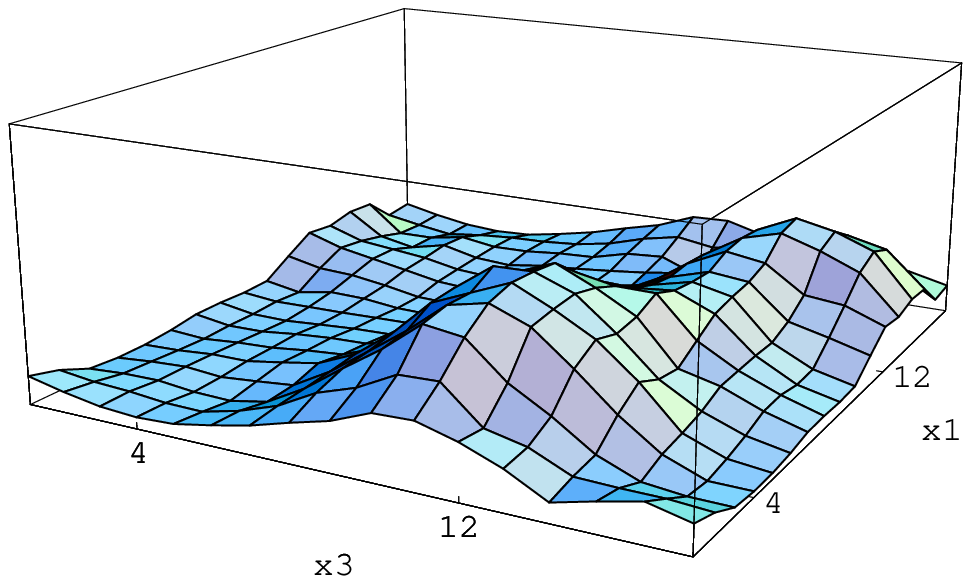}
\includegraphics[width=0.32\linewidth]
{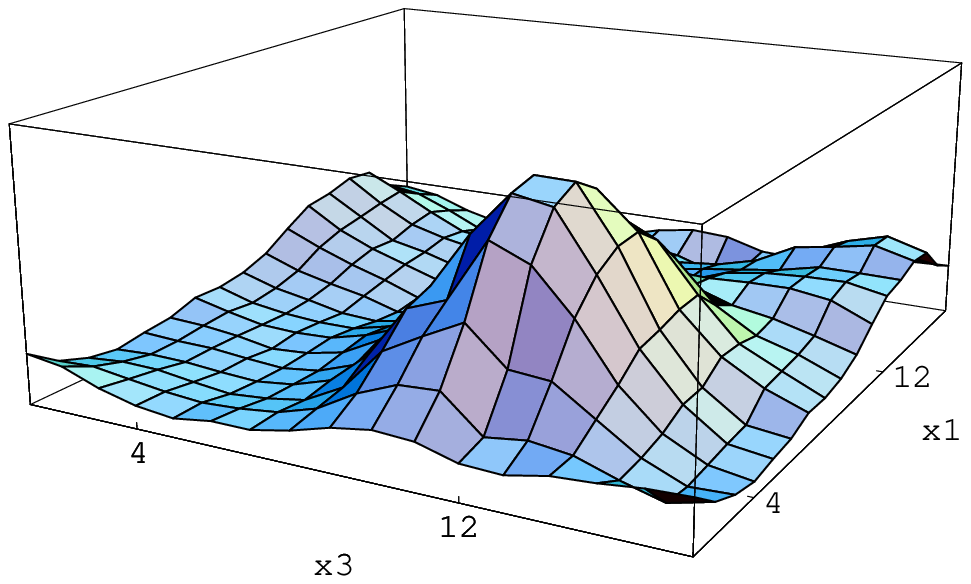}\\
\vspace{0.5cm}

\includegraphics[width=0.32\linewidth]
{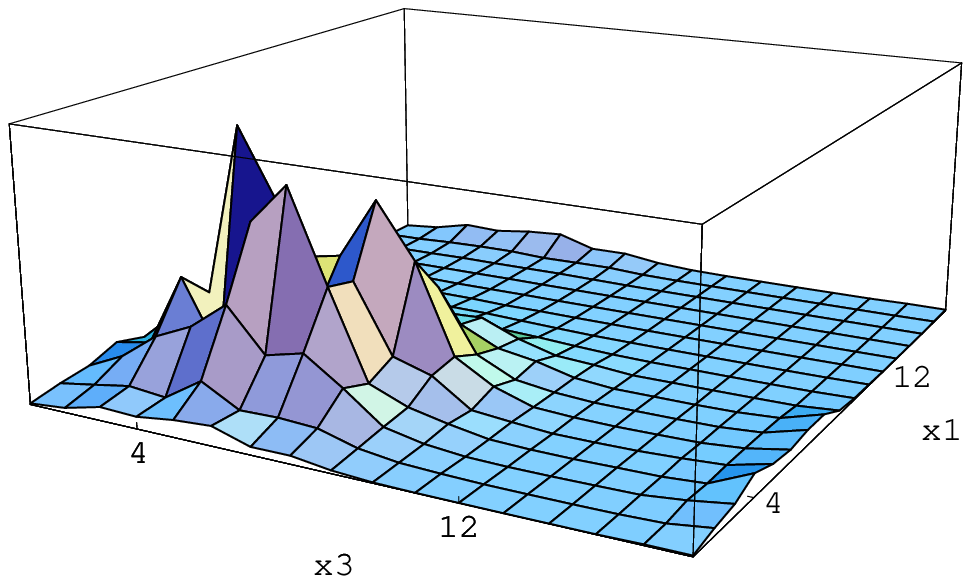}
\includegraphics[width=0.32\linewidth]
{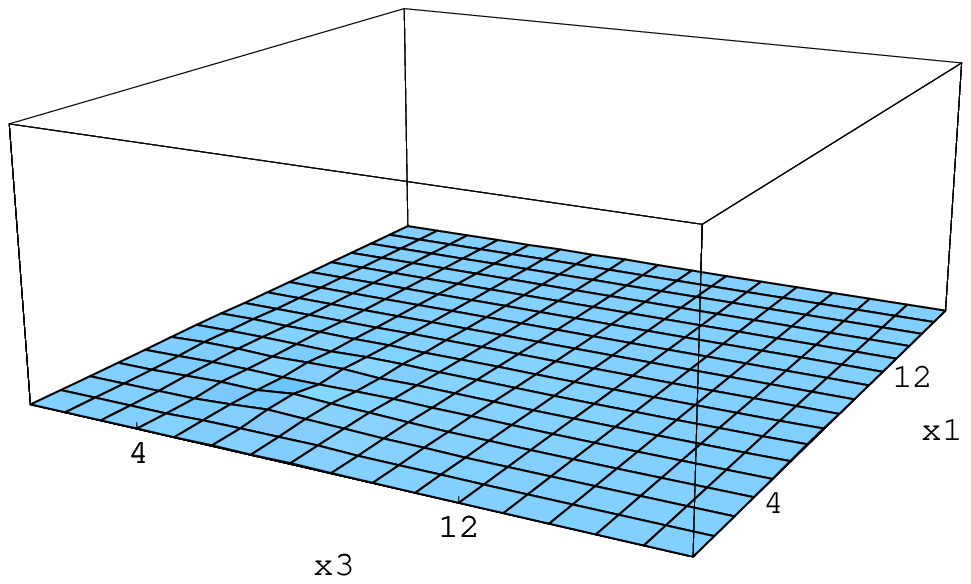}
\includegraphics[width=0.32\linewidth]
{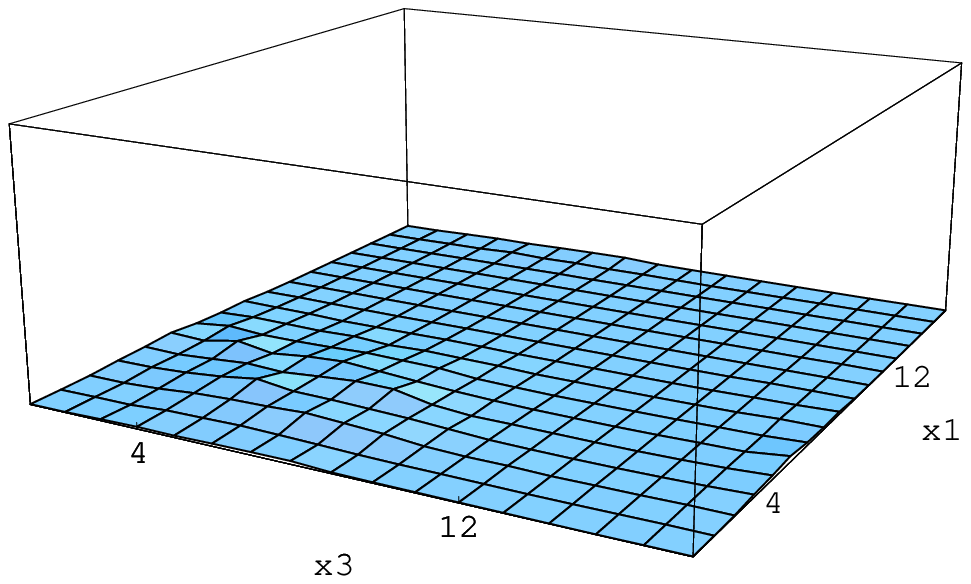}\\
\includegraphics[width=0.32\linewidth]
{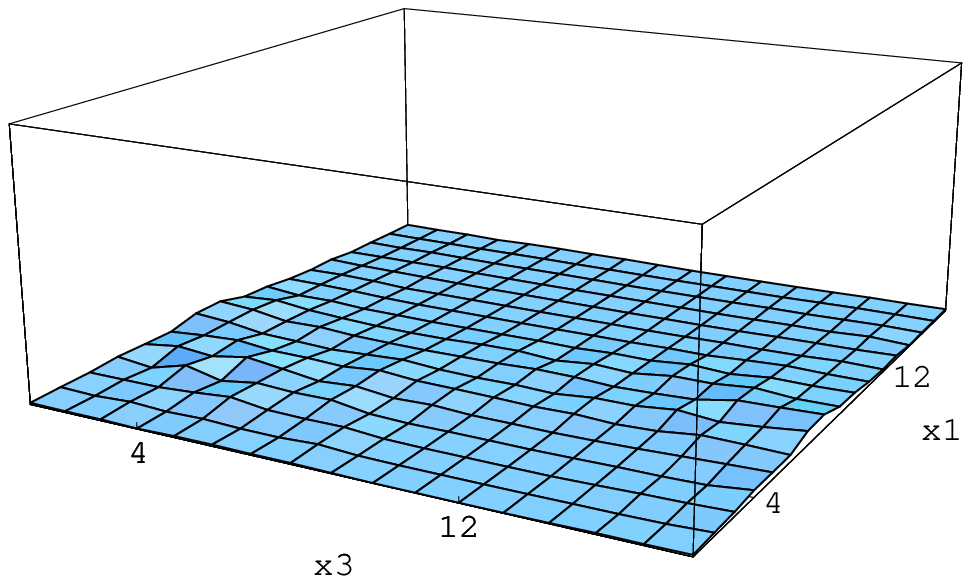}
\includegraphics[width=0.32\linewidth]
{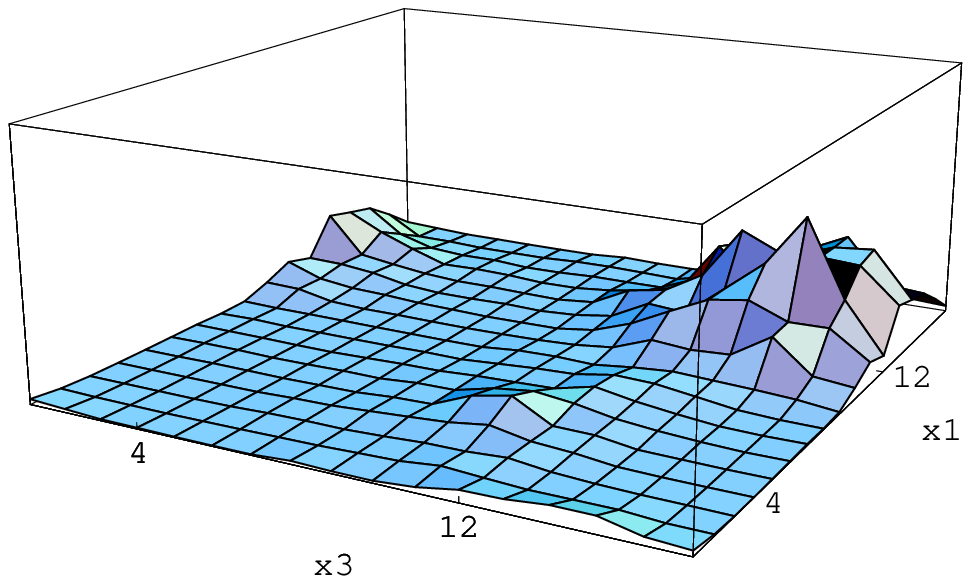}
\includegraphics[width=0.32\linewidth]
{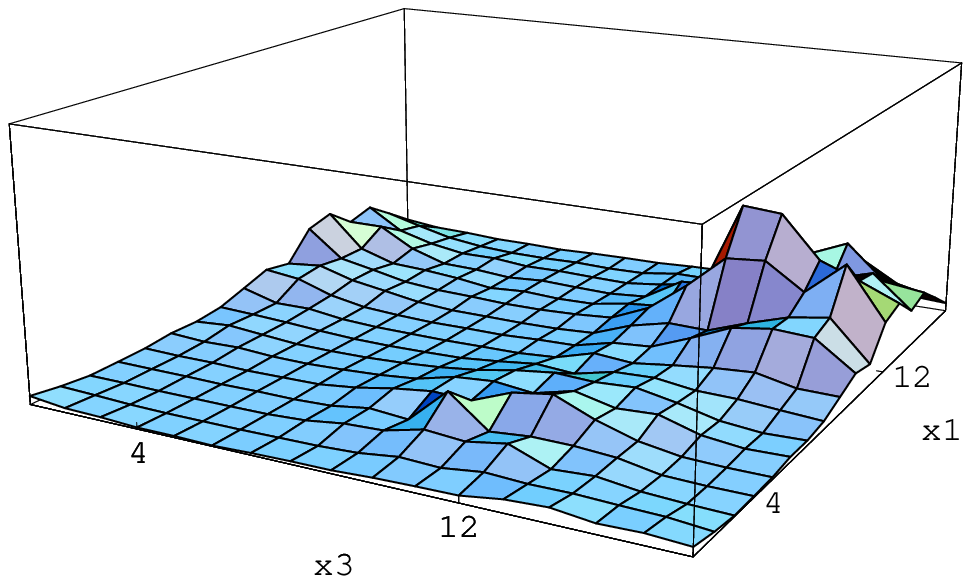}\\
\includegraphics[width=0.32\linewidth]
{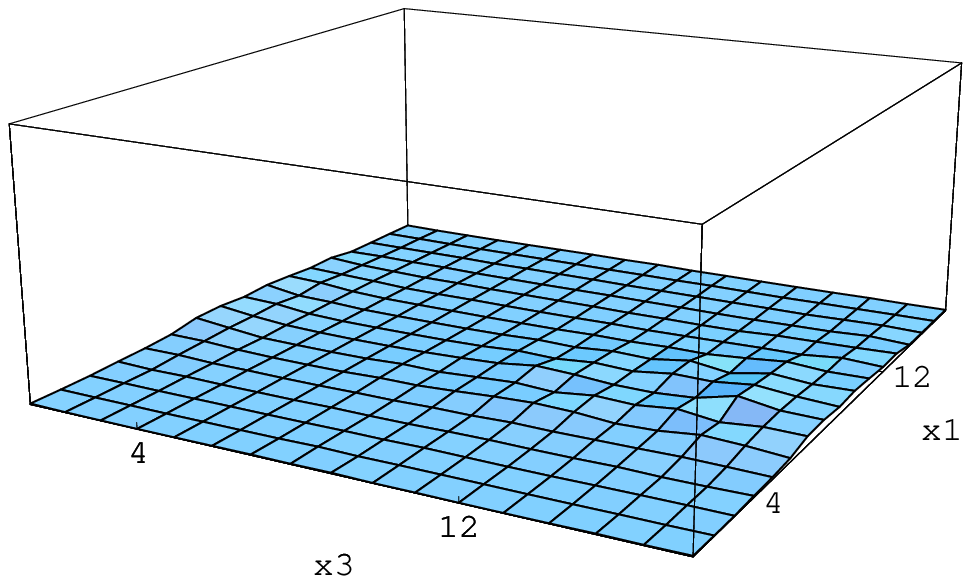}
\includegraphics[width=0.32\linewidth]
{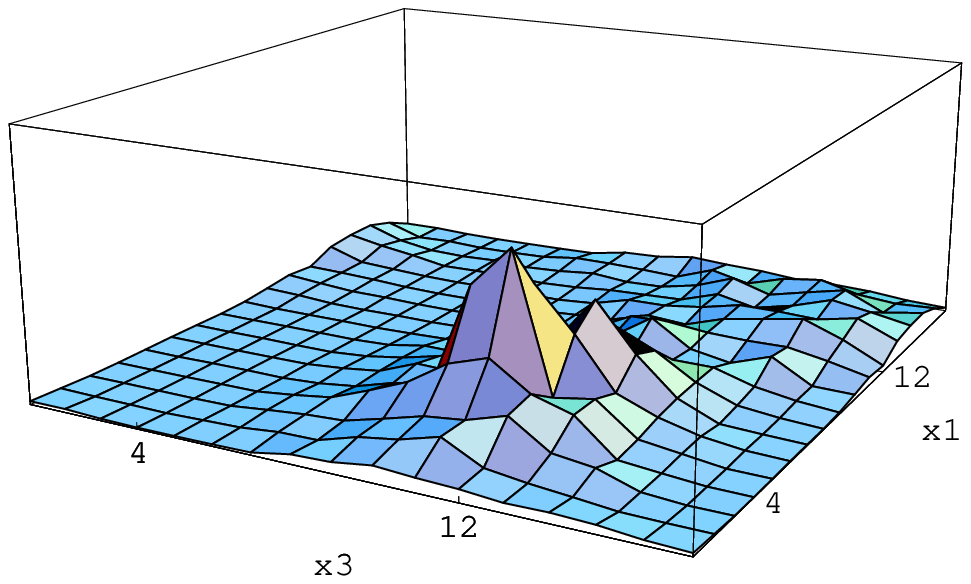}
\includegraphics[width=0.32\linewidth]
{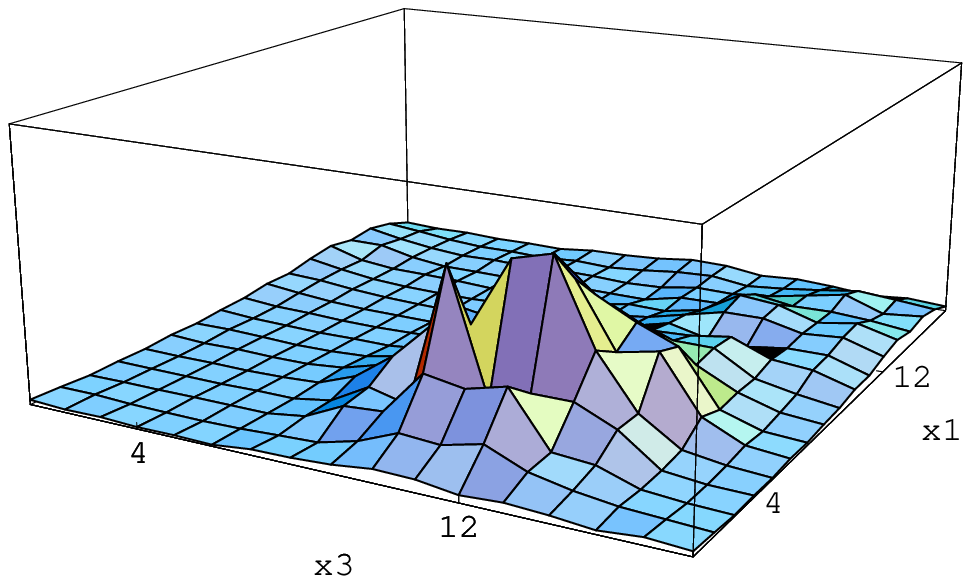}
\caption{First three rows: the lowest Laplacian mode in a thermalized background with
%  boundary condition 
$\zeta=0.02$, $0.355$, $0.48$ from left to right and in
lattice planes $(x_2,x_4)=(12,1),(4,4),(6,3)$ from first to third row
(vertical scale 0.05).
The last three rows show the corresponding profiles of the 
kinetic term (vertical scale 0.0025).}
\label{fig_hop}
\end{figure*}

In this subsection we explore the Laplacian modes as an analyzing  tool on
thermalized gauge field configurations \cite{bruckmann:05b}.
Fig.\ \ref{fig_hop} shows the
lowest-lying mode in a generic background obtained on a $16^3\cdot 4$
lattice at $\beta=2.2$ (which amounts to $T=0.75\, T_c$) confirming
the expectation that these modes are free of
UV fluctuations.
The `evolution' of the Laplacian mode
with the angle $\zeta$ is shown horizontally in Fig.\ \ref{fig_hop}, 
while the first three rows
represent different but fixed lattice planes, which contain the global
maximum of the mode. In the example we find three lattice locations which
become the global maximum within some $\zeta$-interval. Therefore, the
lowest Laplacian mode in a thermalized background hops, too. 

At the hopping  points in $\zeta$ the value of the
maximum\footnote{
The global minimum is not stable enough to employ it for analyzing purposes.}
as well as the inverse participation ratio 
(taking values between ${\rm IPR}\!=\!4$ and ${\rm IPR}\!=\!11$) 
decrease\footnote{
The $\zeta$-values of Fig.\ \ref{fig_hop} are chosen such that the IPR is maximal
within the corresponding $\zeta$-interval.} and the mode rearranges itself.
For some boundary conditions the lowest Laplacian mode can be
characterized as a global structure \cite{bruckmann:05b}.
As the three last rows of Fig.\ \ref{fig_hop} show, the kinetic term follows the
modulus of the mode, but is less smooth.

\begin{figure}[t]
\includegraphics[width=0.7\linewidth,angle=270]
{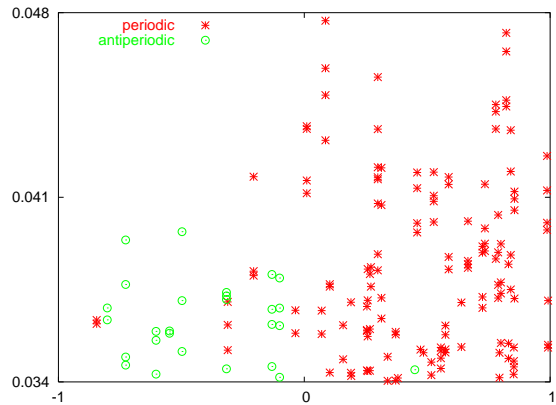}
\caption{Scatterplots of the Polyakov loop (horizontal) at lattice
sites where the modulus of the lowest Laplacian mode (vertical) is large:
stars represent the periodic mode, circles the antiperiodic one.}
\label{fig_mode_vs_polloop}
\end{figure}

Analyzing 50 independent configurations we found that the lowest mode 
hops up to four times as a function of $\zeta$. 
Apart from the
Laplacian mode being wave-like,
the hopping phenomenon resembles the behavior of
the fermion zero mode in thermalized configurations \cite{gattringer:02b}.
However, in both cases it is not obvious which gluonic feature,
e.g.\ in the topological charge,
distinguishes the preferred locations.
Interestingly, we have found a correlation of the maximum
of the periodic and antiperiodic Laplacian mode
to positive and negative Polyakov loops, respectively, 
see Fig.\ \ref{fig_mode_vs_polloop}.
This is the same tendency as for calorons (cf.\ Fig.\ \ref{fig_cal_scal_mode})
which could be understood as an argument
in favor of calorons `underlying' the quantum configurations. 
Alternatively, the Polyakov
loop could play a role defining pinning centers for the Laplacian
mode in the spirit of Anderson localization in a random
potential \cite{anderson:58}.

The lowest Laplacian mode in the {\em adjoint} representation we have
found to have {\em minima} at the caloron monopoles (see also
\cite{bruckmann:01a,alexandrou:00a}) 
that extend to two-dimensional sheets for antiperiodic
boundary conditions \cite{bruckmann:05b}.
In thermalized backgrounds, the maxima of the
adjoint modes are
correlated to the fundamental ones with the same boundary condition,
but are stronger localized than the latter.

\section{A Fourier-like filter}

Laplacian (`harmonic') eigenmodes can be used to 
define a new Fourier-like filter
which we present now \cite{bruckmann:05b}. 
We were inspired by the representation of the
field strength in terms of fermionic modes in \cite{gattringer:02c}, 
but here we aim to reconstruct directly the link variables on which
any observable can be measured. To this
end we combine the definition of the lattice Laplacian, Eq.\ (\ref{eqn_1}), with a
spectral decomposition and at $y=x+\hat{\mu}$ immediately obtain 
\begin{eqnarray}
U_\mu^{ab}(x)&\!\!\!\!=
&\!\!\!\!-\sum_{n=1}^\mathcal{N}\lambda_n\phi^a_n(x)\phi^{*b}_n(x+\hat{\mu})\,,
\nonumber\\
&&\mathcal{N}=N_1N_2N_3N_4\cdot 2\,.
\label{eqn_2} 
\end{eqnarray}
The idea is to truncate the sum on the right hand side 
of this equation at a small mode number $N$ compared to their
total number $\mathcal{N}$.

The question arises, how to relate such an expression to a
{\em unitary} link variable. 
Here the charge conjugation helps, since it guarantees that
every eigenvalue is two-fold degenerate with eigenfunctions related as
$\phi'^a(x)=\epsilon^{ab}\phi^{*b}(x)$. The corresponding
bilinear expressions in Eq.\ (\ref{eqn_2}) add up to an element of
$SU(2)$ up to a factor. 
Such a situation is familiar from the staple average in cooling
or smearing.
We divide by the square root of this factor, which is the determinant and
positive for all practical purposes, and obtain the
final filter formula
\begin{eqnarray}
&&\tilde{U}_\mu^{ab}(x)_N=
(-\sum_{n=1}^N\lambda_n[\phi^a_n(x)\phi^{*b}_n(x+\hat{\mu})-\nonumber\\
&&\epsilon^{ac}\phi^{*c}_n(x)\phi^d_n(x+\hat{\mu})\epsilon^{db}\,]
)/\sqrt{\det (\ldots)}\,.
\label{eqn_3} 
\end{eqnarray}

The quality of the filter is controlled by $N$, where $N\!\!=\!\cal{N}$
reproduces the original configuration exactly. 
%(with no determinant correction necessary) 
 In the other extreme case, \mbox{$N\!=\!1$}, the filtered
links $\tilde{U}_\mu(x)$ can be shown to be pure gauge.
From the behavior of $\phi(x)$ under gauge transformations it
follows immediately that $\tilde{U}_\mu(x)$ transforms as a
link (and no gauge fixing is involved in the filter).

The filtering can be performed with Laplacian modes of any boundary
condition. Then Eq.\ (\ref{eqn_3}) is just the average over opposite
boundary conditions $\zeta$ and $-\zeta$. This additional parameter $\zeta$
completely fixes the Polyakov loop at $N=1$ to 
${\rm tr}\,\tilde{\mathcal{P}}(\vec{x})/2=\cos(2\pi\zeta)$, 
while for nontrivial cases $N\geq 2$ 
the Polyakov loop is observed to fluctuate around this value. It
follows that in order to optimize the filter for our circumstances
(confined phase, nontrivial holonomy calorons) the best choice is
the intermediate case $\zeta=1/4$.

\begin{figure}[!h]
\centering
\includegraphics[width=0.8\linewidth]
{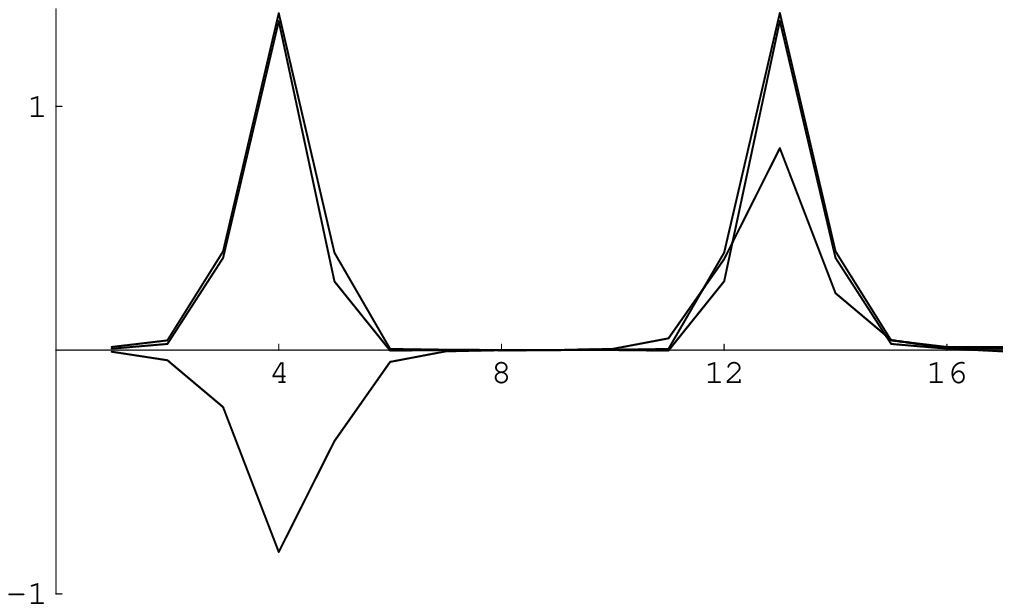}
\includegraphics[width=0.8\linewidth]
{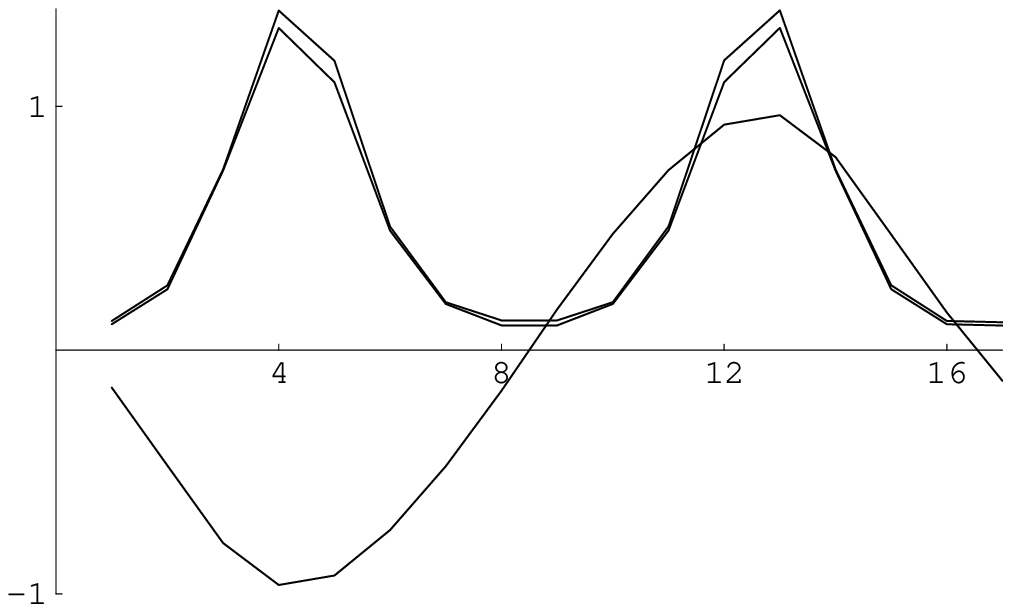}
\includegraphics[width=0.8\linewidth]
{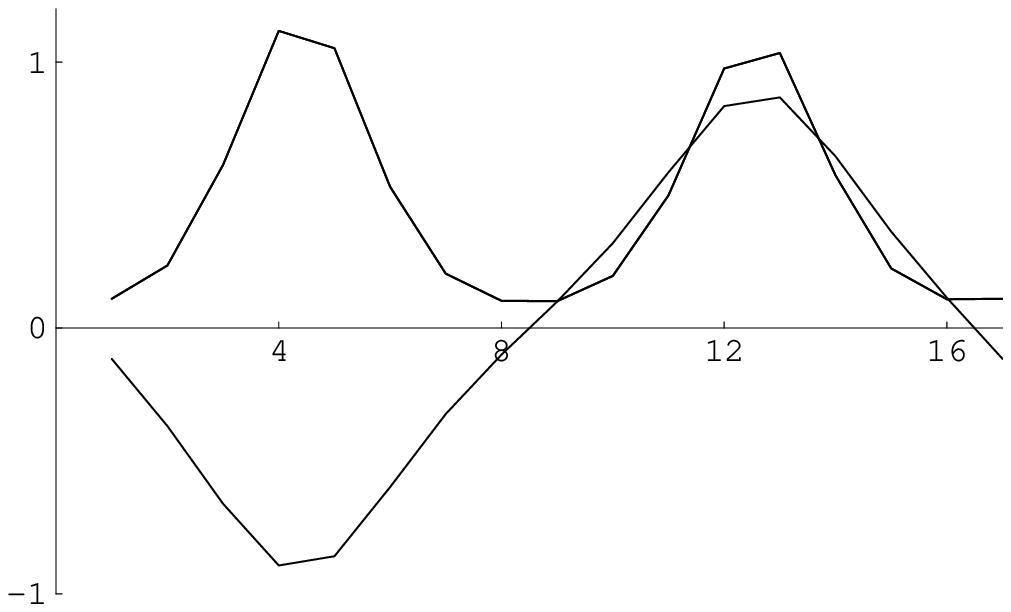}
\caption{Action and topological density and Polyakov loop of a large caloron
on $16^3\cdot 4$,
filtered with boundary condition $\zeta=1/4$.
From top to bottom: $N=4$ modes (densities multiplied by  a factor of 100),
$N=150$ modes (factor 400) and the original configuration
(factor 400).}
\label{fig_cal_filtered}
\end{figure}

In order to test the filter, we again start with a caloron
(this time over $16^3\cdot 4$).
The action and topological density as well as the Polyakov loop
plotted in Fig.\ \ref{fig_cal_filtered} show, that the filter
with the number of modes as low as $N=4$
starts to reproduce the classical structures qualitatively, while for
$N=150$ the agreement is almost perfect.

During the filtering we have recorded the determinant used in Eq.\
(\ref{eqn_3}) to project to an $SU(2)$-element. With increasing number $N$ of
modes it increases, too (i.e.\ becomes closer to 1). Moreover, its variation
over the lattice decreases, such that all links appear `equally well filtered'.\\

Now we turn to the application of the filter to thermalized configurations (at
finite temperature), first by the example %configuration
of Sect.\ \ref{sect_2.2}.

\begin{figure*}
\includegraphics[width=0.33\linewidth]
{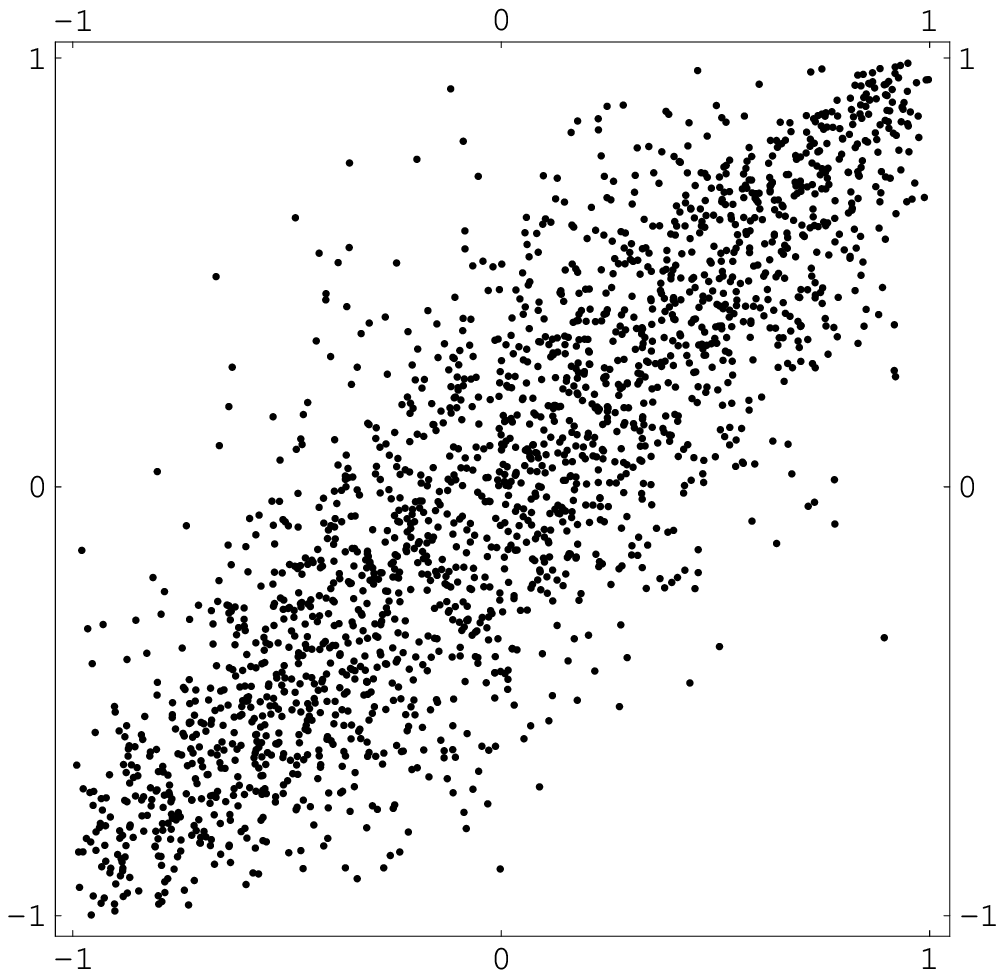}
\includegraphics[width=0.33\linewidth]
{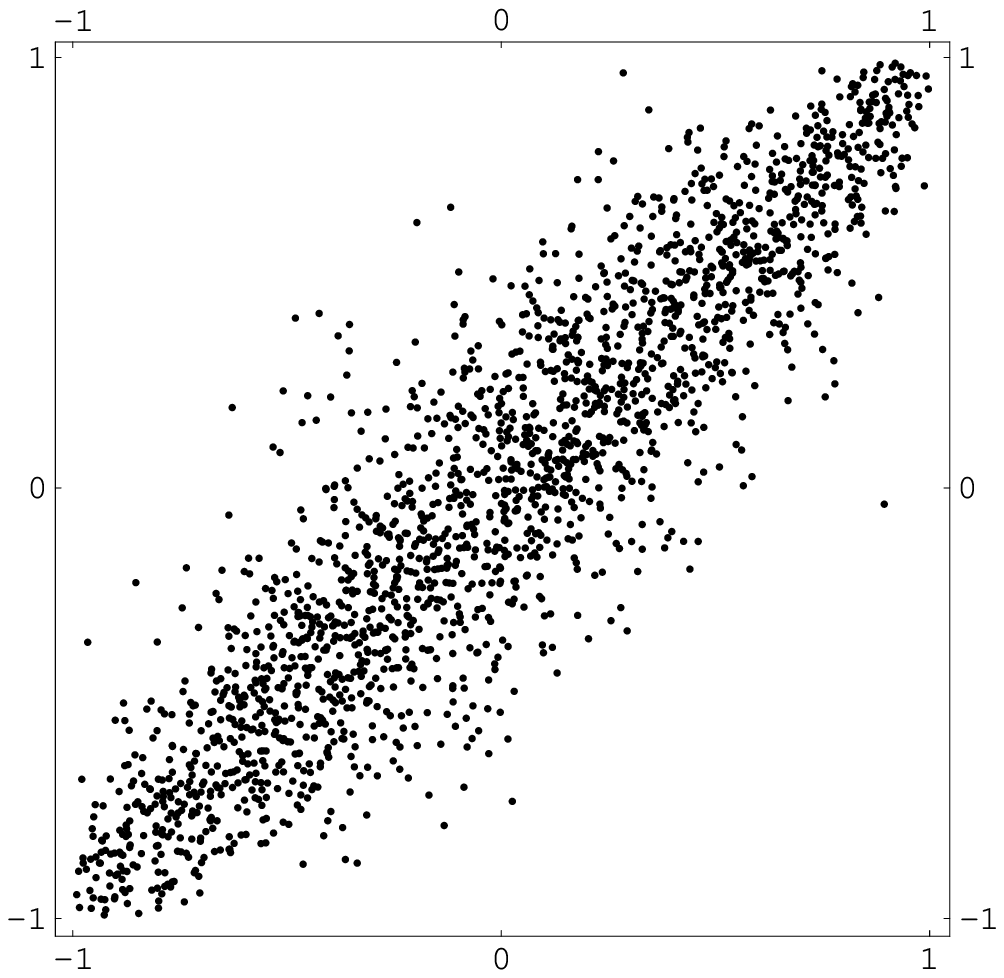}
\includegraphics[width=0.33\linewidth]
{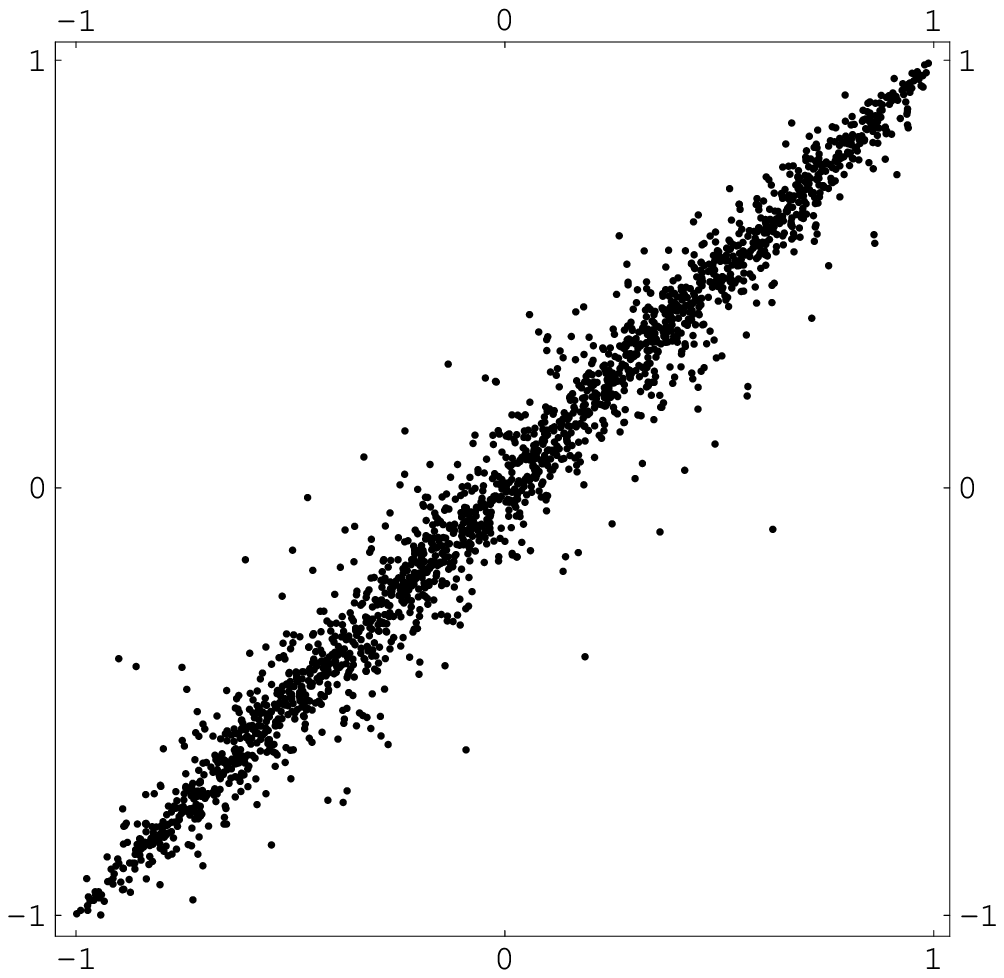}
\caption{Correlation of link variable components
${\rm tr}\, U_\mu(x)/2$ and ${\rm tr}\, U_\mu(x)\tau_i/2i$
($\tau_i$ the Pauli matrices) between the original configuration and
the filtered one with $N=2$ (left) and $N=100$ (middle) and 
within filtered links with $N=2$ vs.\ $N=4$ (right). In order not to overload
the plots, only 2000 link variable components have been plotted.}
\label{fig_corr_links}
\end{figure*}

The first question we want to address is the correlation of the filtered links
to the original ones.
Fig.\ \ref{fig_corr_links} shows scatterplots including all $SU(2)$-components of the
links in all $\mu$-directions and over all lattice locations. To
compare link variables themselves makes sense, since the original and the
filtered configuration come in the same gauge (in other words the application of
the filter commutes with gauge transformations).
Fig.~\ref{fig_corr_links} (left) shows that even in the first nontrivial case $N=2$ the
filtered links are correlated to the original ones. With increasing number of
involved modes this correlation becomes stronger as expected:
in the scatterplot for $N=100$ (Fig.~\ref{fig_corr_links} (middle)) 
the deviation from the
diagonal decreased compared to $N=2$. One can make this statement quantitative
by computing the average square distance from the diagonal, which indeed is
smaller for bigger $N$. Actually, it also shows that the time-like links are
less correlated than the space-like ones, especially
when using the `inappropriate' periodic boundary conditions.
Finally, Fig.\ \ref{fig_corr_links} (right) shows 
that the filter with comparable number of
modes produces strongly correlated links.\\

\begin{figure*}
\centering
\includegraphics[width=0.35\linewidth,angle=270]
  {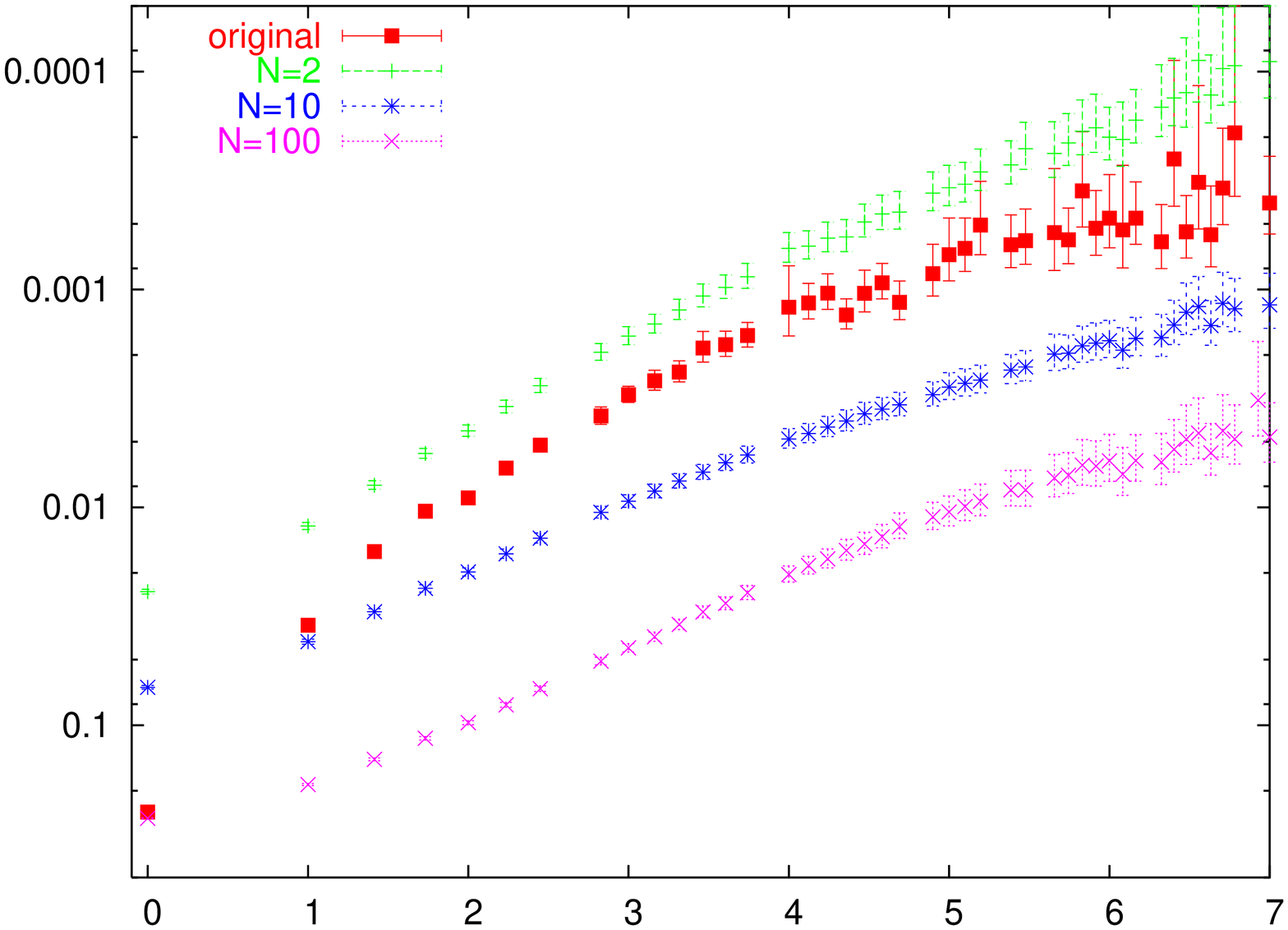}
\includegraphics[width=0.35\linewidth,angle=270]
  {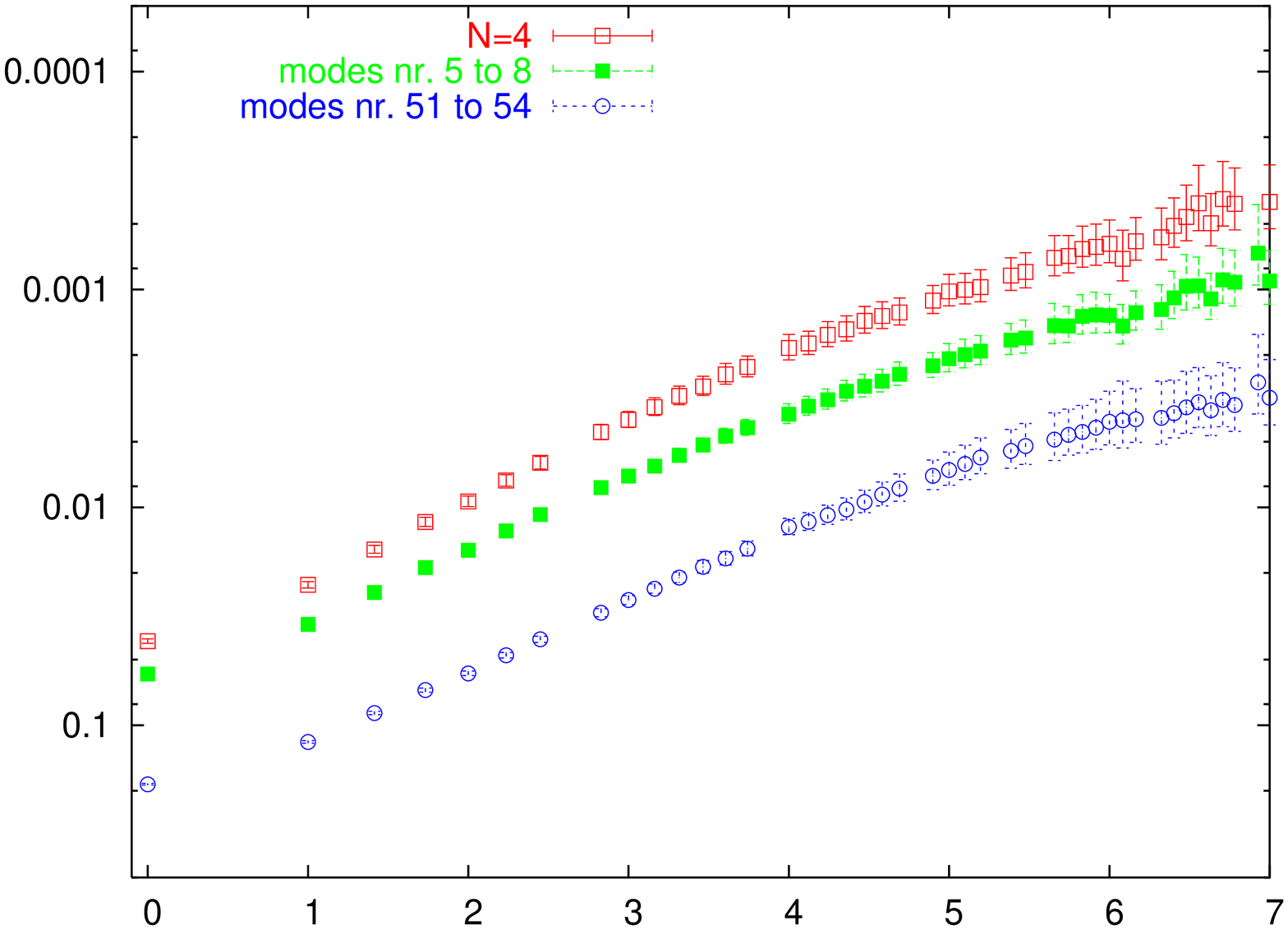}
\caption{Polyakov loop correlators of 50 configurations at finite
temperature, filtered with boundary condition $\zeta=1/4$ and
plotted on an inverse logarithmic scale over the distance. 
Left: the original configurations vs.\ the filtered ones with
$N=2,\,10,\,100$ modes. Right: 4 modes with various shifts in the
ordinal number of the modes.}
\label{fig_stringtensions}
\end{figure*}

What is even more remarkable is that the filter method preserves the string
tension. We plot the logarithm of the Polyakov loop correlator
measured on 50 configurations at $T=0.75\,T_c$ 
in Fig.\ \ref{fig_stringtensions} (left),
both for the original one and the filtered ones with $N=2,\,10,\,100$.
It reveals a clearly linear behavior with a slope
reproducing the original one within 15\%. 
The minimal number of modes, $N=2$, suffices for this behavior.
We conclude that the
confining properties of lattice gauge theory are captured by the
lowest Laplacian modes.

This property actually does not depend much on the ordinal number of the modes,
see Fig.\ \ref{fig_stringtensions} (right). 
First of all, one can include the $N$ highest
Laplacian modes. Due to the symmetry \cite{greensite:05}
\begin{equation}
\phi_{\mathcal{N}-n}=(-1)^{\sum_\mu x_\mu}\phi_{n}\,,\quad 
\lambda_{\mathcal{N}-n}=16-\lambda_{n}\,,
\end{equation}
including these highest modes results in the same bilinear
contributions in Eq.\ (\ref{eqn_3})
with the only modification $\lambda_n\rightarrow 16-2\lambda_n$.
The Polyakov loop correlator computed from the 50 configurations filtered in
this way agrees very well 
with the one discussed so far (and is therefore not
shown in Fig.\ \ref{fig_stringtensions}). 
The second possibility is to shift the mode number
from 1 till 4 to say 5 till 8 or 51 till 54.
The corresponding curves shown 
in Fig.\ \ref{fig_stringtensions} (right) reveal the same linear
behavior up to a vertical shift.
The Polyakov loop correlator is robust also against a third option, namely
replacing the eigenvalues $\lambda_n$ in Eq.\ (\ref{eqn_3}) by a
constant,
which again produces identical curves (not shown).

\begin{figure}
\centering
\includegraphics[width=0.6\linewidth]
{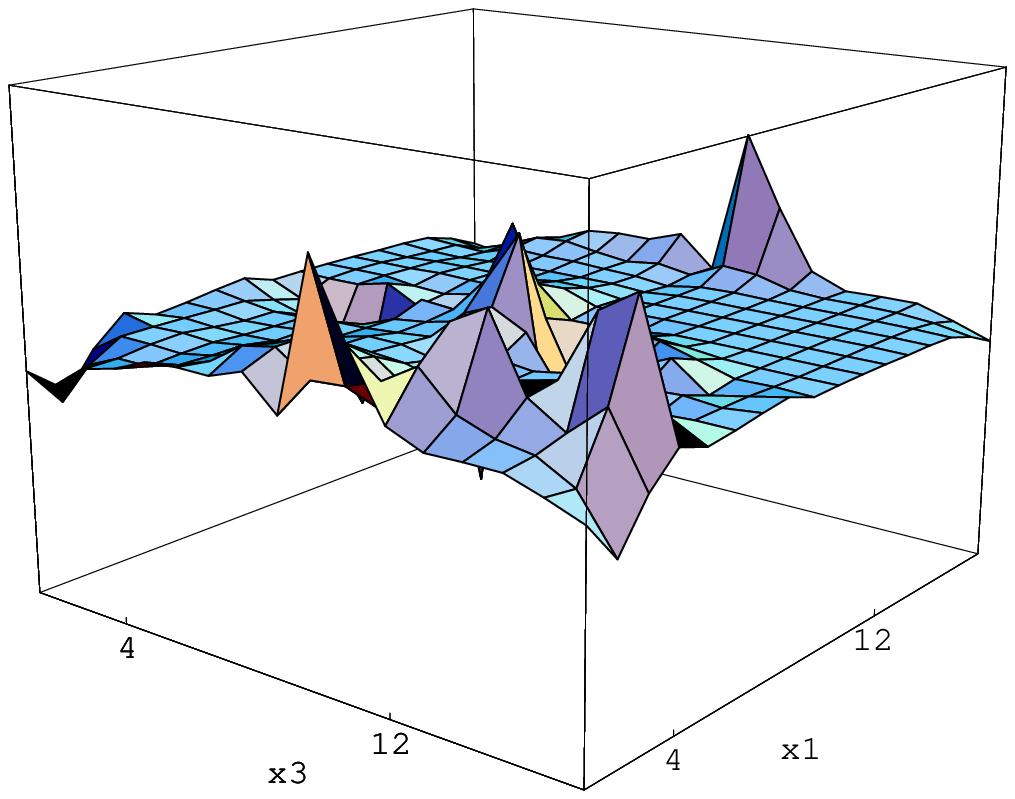}
\vspace{0.3cm}

\includegraphics[width=0.6\linewidth]
{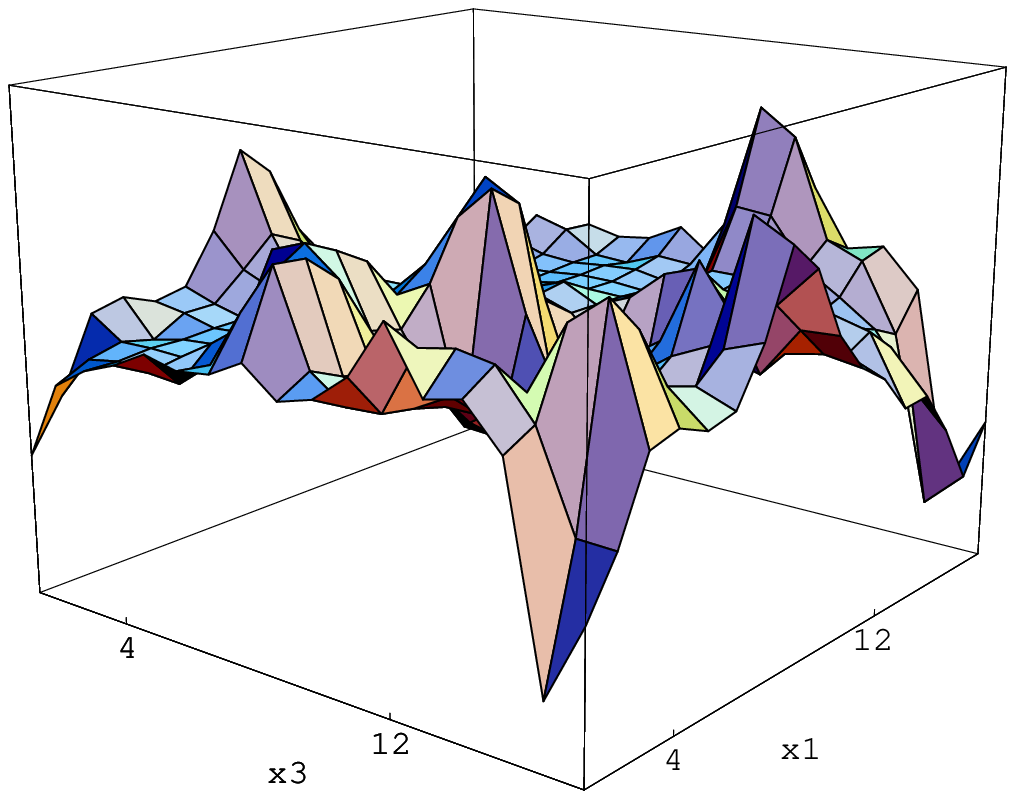}
\vspace{0.3cm}

\includegraphics[width=0.6\linewidth]
{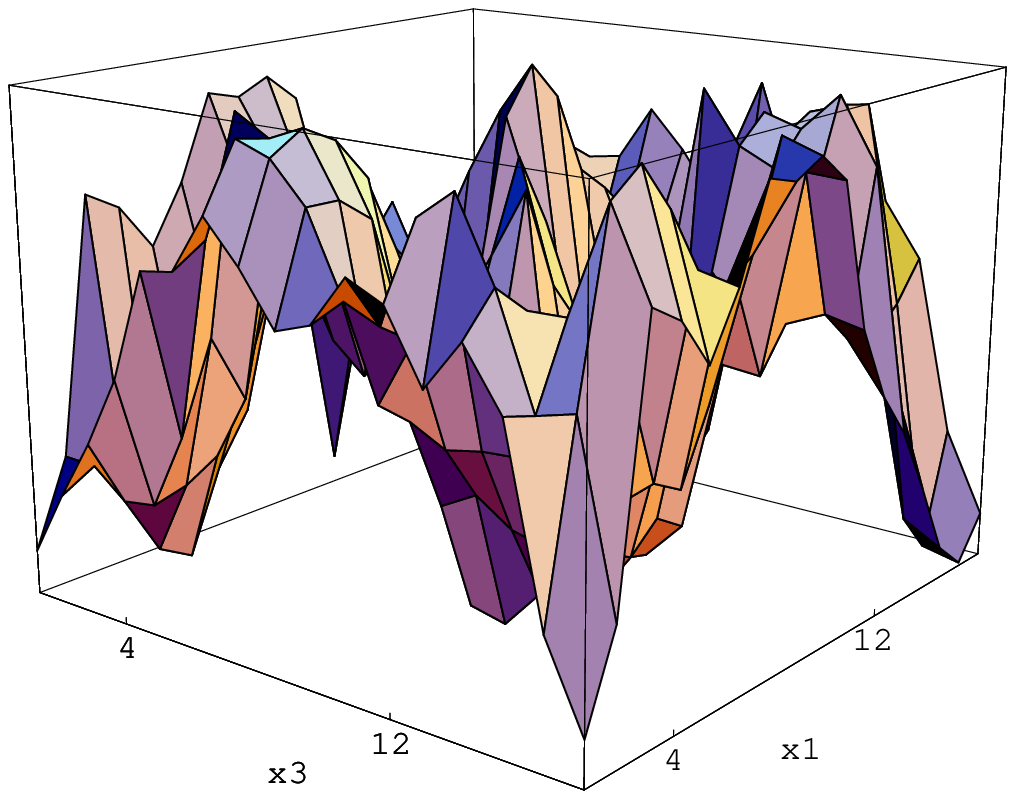}
\vspace{0.3cm}

\includegraphics[width=0.6\linewidth]
{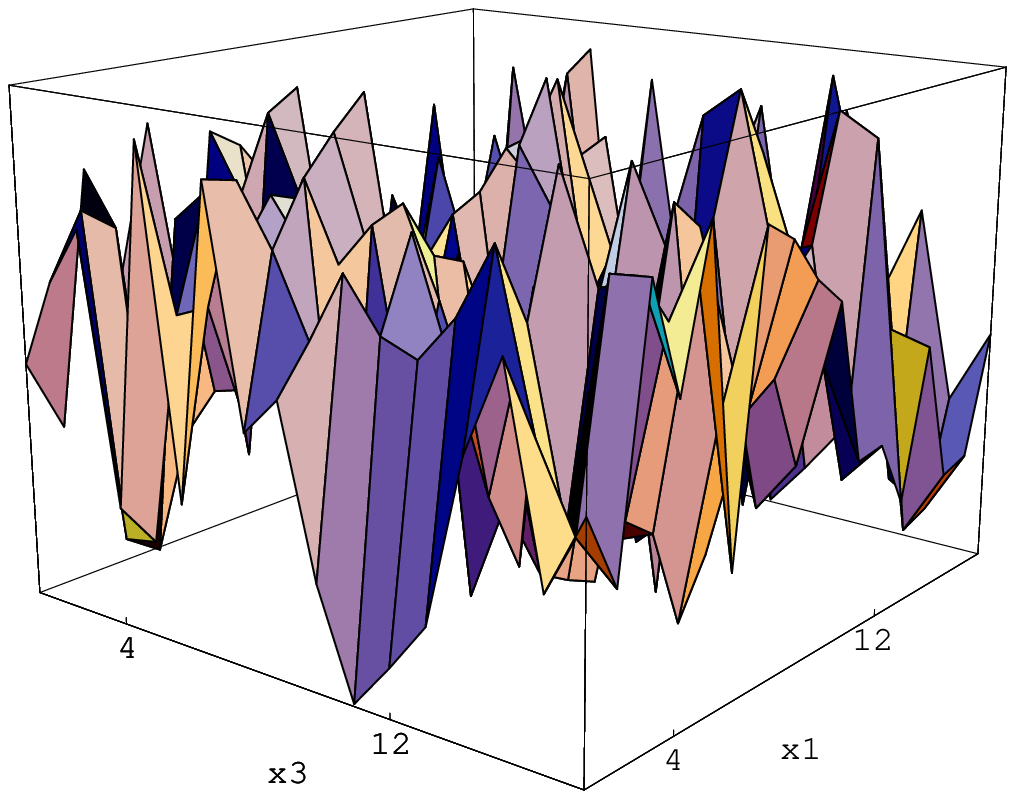}
\caption{Polyakov loop `evolution' with increasing number of modes for
the thermalized example in a fixed lattice plane ($x_2=6$). From top to
bottom: $N=2,\,10,\,100$ and the original configuration. The
vertical axes are from $-1$ to $1$.}
\label{fig_polloops_filtered}
\end{figure}

After filtering the Polyakov loop correlator has no sign of a Coulomb
regime, since the filter has washed out short range fluctuations. At
zero distance the values 
$\langle({\rm tr}\,\tilde{\mathcal{P}}(\vec{x}))^2\rangle$
are smaller than the original one.
This reflects the already mentioned effect that the filtered links for small
$N$ are mostly traceless (governed by the choice $\zeta=1/4$). The
site distribution of the Polyakov loop is a narrow one around 
${\rm tr}\,\tilde{\mathcal{P}}=0$ for
$N=2$, but for $N=100$ is compatible with the original distribution
which in turn is described by the Haar measure. 
Fig.\ \ref{fig_polloops_filtered} shows locally (for the example
configuration and in a fixed lattice plane) that with growing $N$ more
and more fluctuations appear in the Polyakov loop. Studies are under way to
clarify whether the new filter method preserves other infrared features, like
the spatial Wilson loops, and how it behaves at zero temperature.\\

Since the filter preserves the string tension from Polyakov loop
correlators and is not biased to classical solutions nor to
particular degrees of freedom in the gauge field 
(like monopoles or vortices), it is
interesting to have a closer look at the vacuum structures that emerge when
the filter is applied to generic configurations.

Concerning the action density, these objects are found to be
isolated peaks (which are non-static and not necessarily (anti)selfdual).
This phenomenon seems counterintuitive as the filter uses the lowest-lying
Laplacian modes which are smooth.
However, too small action density lumps have also been observed when filtering
smooth calorons at low $N$ (cf.\ Fig.\ \ref{fig_cal_filtered} (top)).
This phenomenon presumably comes from the nonlinearity in the filter formalism
(Eq.\ (\ref{eqn_3})), in particular the determinant used to scale up the mode
contributions to an $SU(2)$-element.
Interestingly, we have found a correlation of the peak structure of the
filtered action density to that
emerging after cooling or smearing in an early stage.

We stress that, in contrast to e.g.\ cooling, the filter does not involve the
plaquette action and thus is not made to reduce it in the first place.
Upon filtering the links, the total action does become smaller, but only down to some
amount (in our example down to 66 instanton units at $N=2$), which might well
be the minimal content necessary to reproduce the infrared physics.
More work has to be done to better understand the 
spiky structures induced by the filter.

\section*{Acknowledgments}

FB likes to thank the organizers for inviting him to a stimulating workshop.
He and EMI thank A.~di Giacomo, S.~D\"urr, J.~Greensite, \v S.~Olejn\'\i k
and U.-J.~Wiese for helpful discussions.

%\bibliographystyle{../JHEP}
%\bibliography{../gauge}

\providecommand{\href}[2]{#2}\begingroup\raggedright\endgroup

\end{document}